\documentclass{ifacconf}

\usepackage{amsmath}
\usepackage{amssymb}

\usepackage{etoolbox}
\let\classAND\AND
\let\AND\relax
\usepackage{algorithm} 
\usepackage{algorithmic}

\let\AND\classAND
\AtBeginEnvironment{algorithmic}{\let\AND\algoAND}

\usepackage{graphicx}      
\usepackage{subfigure}
\graphicspath{{figures/}} 
\usepackage{bm}

\makeatletter
\let\old@ssect\@ssect 
\makeatother

\usepackage{natbib}
\usepackage{hyperref}

\makeatletter
\def\@ssect#1#2#3#4#5#6{%
  \NR@gettitle{#6}
  \old@ssect{#1}{#2}{#3}{#4}{#5}{#6}
}
\makeatother

\allowdisplaybreaks[4]

\newcommand{\RR}{\mathbb{R}}

\newcommand{\NN}{\mathbb{N}}
\newcommand{\ZZ}{\mathbb{Z}}
\newcommand{\eps}{\varepsilon}

\newcommand{\ModelDeGroot}{\mathcal{M}_{\tx{DG}}} 
\newcommand{\ModelFJ}{\mathcal{M}_{\tx{FJ}}}
\newcommand{\ModelRepell}{\mathcal{M}_{\tx{RP}}} 
\newcommand{\ModelHK}{\mathcal
{M}_{\tx{HK}}} 

\newcommand{\SetAllModels}{\{\ModelDeGroot, \ModelFJ, \ModelRepell, \ModelHK\}} 

\newcommand{\ModelFori}{\mathcal
{M}_{i}} %

\newcommand{\mtce}{\mathcal{E}}

\newcommand{\mtcg}{\mathcal{G}}

\newcommand{\mtci}{\mathcal{I}}

\newcommand{\mtcn}{\mathcal{N}}

\newcommand{\mtcv}{\mathcal{V}}

\newcommand{\Let}{: =}

\newcommand{\sgn}{\textup{sgn}}

\newcommand{\tp}{\text{T}}

\newcommand{\bfl}{\mathbf{1}}

\newcommand{\tx}{\textup}

\begin{document}
\begin{frontmatter}

\title{Joint Learning of \\Network Topology and Opinion Dynamics \\Based on Bandit Algorithms\thanksref{footnoteinfo}} 

\thanks[footnoteinfo]{This work was supported by the Knut \& Alice Wallenberg Foundation, the Swedish Research Council, and the Swedish Foundation for Strategic Research.}

\author[kth]{Yu Xing} 
\author[kth]{Xudong Sun}
\author[kth]{Karl H. Johansson} 

\address[kth]{ Division of Decision and Control Systems, EECS, KTH Royal Institute of~Technology; Digital Futures,
SE-10044 Stockholm, Sweden \\(e-mail: yuxing2@kth.se, smilesun.east@gmail.com, kallej@kth.se).}


\begin{abstract}    
We study joint learning of network topology and a mixed opinion dynamics, in which agents may have different update rules. Such a model captures the diversity of real individual interactions. We propose a learning algorithm based on multi-armed bandit algorithms to address the problem. The goal of the algorithm is to find each agent's update rule from several candidate rules and to learn the underlying network. At each iteration, the algorithm assumes that  each agent has one of the updated rules and then modifies network estimates to reduce validation error. Numerical experiments show that the proposed algorithm improves initial estimates of the network and update rules, decreases prediction error, and performs better than other methods such as sparse linear regression and Gaussian process regression.  
\end{abstract}


\end{frontmatter}

\section{Introduction}
Opinion dynamics characterize how individuals interact with each other and change their opinions, which has attracted researchers in various disciplines from control to physics for decades (\cite{
proskurnikov2017tutorial
}). 
There is a growing interest in learning such dynamics (\cite{ravazzi2021learning}). Most researches formulate the learning problem based on a single  model. In real networks agents may have different update rules
, so there is a need to study how to learn such networked dynamics.

\subsection{Related Work}\label{sec:related}
This paper considers opinion dynamics with continuous states 
(\cite{proskurnikov2017tutorial}). A classic example is the DeGroot model (\cite{degroot1974reaching}), in which agents update to the opinion average of their neighbors. 
\cite{friedkin1999social} generalize this model by assuming the agents are influenced by their initial positions. 
Bounded confidence models, such as the Hegselmann–Krause (HK) model (\cite{hegselmann2002opinion})
, characterize the case when 
agents only interact with those who hold opinions similar to themselves. 
Opinion dynamics over signed networks are another important class of models (\cite{shi2019dynamics}). Signed networks have not only positive but also negative edges. 
There has been research on testing opinion dynamics models against data. \cite{clemm2017micro} find the existence of linear averaging. The FJ model has been validated in small-group experiments (\cite{friedkin1999social
}). Recent empirical studies based on large-scale datasets (\cite{
kozitsin2021opinion,
kozitsin2022formal}) also suggest the presence of bounded confidence and negative influence rules.

With the increasing availability of large-scale online datasets, the interest in learning influence networks has been growing (\cite{ravazzi2021learning}). 
\cite{de2014learning} and~\cite{wai2016active} study sparse network learning of the DeGroot model.
\cite{ravazzi2017learning} investigate the learning of the FJ model, 
and \cite{ravazzi2021ergodic} consider the case where only partial observations are available. \cite{wai2019joint} study joint learning of network topology and system parameters for bounded confidence models. Learning based on quantized observations are studied in~\cite{xing2022recursive,xie2023finite}. 
Most of these papers assume that the type of the dynamics model to be learned is known and every agent follows the same update rule. This is often not the case in real  networks, so it is necessary to study learning of networked dynamics mixing multiple update rules. 
Automated machine learning (AutoML) is an emerging domain studing how to design algorithms that automatically build machine learning (ML) models with optimized hyperparameter settings (\cite{he2021automl}). 
\cite{thornton2012auto} introduce the sequential model-based algorithm configuration
to select algorithms and optimize hyperparameters. \cite{li2017hyperband,sun2019reinbo
} propose methods based on multi-armed bandit and reinforcement learning algorithms.

\subsection{Contribution}
We study joint learning of network topology and a mixed opinion dynamics, where agents may have different update rules. In the learning problem, the network and the types of agent update rules are unknown. We propose a learning algorithm based on multi-armed bandits to address the problem. 
The algorithm starts with initial estimates of the network and the agent  update rules. At each iteration, it either refines the network estimates by exploiting a given update rule by modifying the adjacency matrix, or examines the validation error of other update rules with small probability. 
Numerical experiments show that the proposed algorithm can improve the initial estimates, with better network recovery and smaller prediction error. 
The algorithm provides a search strategy for learning opinion dynamics models with multiple update rules, and performs better than linear sparse regression and Gaussian process regression.
Compared with existing network learning methods, the proposed algorithm can get rid of the assumption of a single model, and can include multiple sets of constraints. Thus the proposed algorithm can provide better possibility for learning real dynamics. The algorithm is inspired by AutoML, but the problem cannot be solved directly by AutoML methods. A multi-agent system is considered here, and the best model needs be selected for each agent. In addition, the network and dynamics are coupled.

\subsection{Outline}
The rest of the paper is organized as follows. Section~\ref{sec_prel} introduces several opinion dynamics models and defines the mixed model. Section~\ref{sec_prob} formulates the learning problem. We propose a learning algorithm for the mixed model in Section~\ref{sec_alg}. Section~\ref{sec_simulation} presents numerical experiments.

\emph{\textbf{Notation.}}
Let $\mathbf{1}_n$ be the $n$-dimensional all-one vector, 
and $\mathbf{0}_{n,m}$ be the $n\times m$-dimensional all-zero matrix. 
The Euclidean norm and the $l_1$ norm are $\|\cdot\|$ and $\|\cdot\|_1$. 
For a vector $x\in \mathbb{R}^n$, denote 
its subvector from $i$-th to $j$-the component by $x_{i:j}$, $i\le j$. For a matrix $A = [a_{ij}] \in \mathbb{R}^{n\times n}$, denote its $(i,j)$-th entry by $a_{ij}$ or $[A]_{ij}$ and denote its $i$-th column (row) by $[A]_{:,i}$ ($[A]_{i,:}$). 
The function $\mathbb{I}_{[\textup{property}]}$ is the indicator function
. If the property is stated for a vector, then the function is entry-wise (e.g., $\mathbb{I}_{[y > 0]} = [\mathbb{I}_{[y_1 > 0]}~\cdots~\mathbb{I}_{[y_n > 0]}]^\tp$ for $y\in \RR^n$). Denote the entry-wise sign of a vector $y$ by $\sgn(y)$. 
The structure of a network is defined by an undirected graph $\mtcg = (\mtcv, \mtce, A)$, where $\mtcv = \{1,\dots,n\}$ is the agent set, $\mtce$ is the edge set, and $A$ is the adjacency matrix. Each edge $\{i,j\} \in \mtce$ has a sign, either positive or negative. The disjoint sets $\mtce^+$ and $\mtce^-$ 
collect all positive and negative edges, respectively. The positive (negative) neighbors of an agent $i$ is denoted by $\mtcn_i^+ = \{j:\{i,j\} \in \mtce^+\}$ ($\mtcn_i^- = \{j:\{i,j\} \in \mtce^-\}$), and $\mtcn_i \Let \mtcn_i^+ \cup \mtcn_i^-$. 
An agent~$i$ has positive and negative degrees $d_i^+ = |\mtcn_i^+|$ and $d_i^- = |\mtcn_i^-|$, respectively, and degree $d_i \Let d_i^+ + d_i^-$. The adjacency matrix $A$ satisfies that $a_{ij} = 1$ if $\{i,j\} \in \mtce^+$, $a_{ij} = -1$ if $\{i,j\} \in \mtce^-$, and $a_{ij} = 0$ if $\{i,j\} \not\in \mtce$. 

\section{Preliminaries}\label{sec_prel}
This section introduces considered models. 
We assume that each agent~$i$ has a self loop $\{i,i\} \in \mtce^+$ 
 and has a state $x_i(t)$ at time $t \in \NN$. Denote the state vector by $x(t) \in \RR^n$.

The DeGroot Model (\cite{degroot1974reaching}) is one of the most classic opinion dynamics. The underlying graph does not have negative edges (i.e., $\mtce^- = \emptyset$). Each agent~$i$ updates its opinion to the weighed average of its neighbors' opinions:
\begin{equation}\label{eq_M1}
    x_i(t+1) = \sum_{j=1}^n w_{ij} x_j(t) =: f^{\ModelDeGroot}(x(t),\theta^{(\ModelDeGroot,i)}), 
\end{equation}
where $t \in \NN$ and $w_{ij}$ is the influence weight of the agent~$j$ on the agent~$i$ and we denote
\begin{equation}\label{eq_theta1}
    \theta^{(\ModelDeGroot,i)} = [w_{i1}~\cdots~w_{in}]^\tp, ~i\in\mtcv.
\end{equation} 
The weight $w_{ij}$ satisfies that $w_{ij} > 0$ if $a_{ij} = 1$, and $w_{ij} = 0$ if $a_{ij} = 0$. In addition, $\sum_{i=1}^n w_{ij} = 1$. 

The FJ model 
generalizes the DeGroot model. Here $\mtce^- = \emptyset$, and every agent~$i$ has susceptibility $\lambda_i \in [0,1]$ to others. An agent~$i$ updates according to the following rule
\begin{align} \nonumber
    x_{i}(t+1) &= \lambda_i \bigg(\sum_{i=1}^n w_{ij} x_j(t) \bigg) + (1 - \lambda_i) x_i(0) \\ \label{eq_M2}
    &=: f^{\ModelFJ}(x(t),\theta^{(\ModelFJ,i)}),
\end{align}
where $w_{ij}$ are nonnegative weights such that $\sum_j w_{ij} = 1$, $w_{ij} > 0$ if $a_{ij} = 1$, and $w_{ij} = 0$ if $a_{ij} = 0$. Here we denote \begin{equation}\label{eq_theta2}
    \theta^{(\ModelFJ,i)} \Let [w_{i1} ~\cdots ~w_{in} ~\lambda_i]^\tp, i\in\mtcv.
\end{equation}
Note that $f^{\ModelFJ}$ also depends on $x_i(0)$, which is omitted for notation simplicity.
When $\lambda_i \not= 1$, the opinion of the agent~$i$ is constantly influenced by its initial position $x_i(0)$. An agent~$i$ with $\lambda_i = 0$ is called stubborn and never changes its opinion. When $\lambda_i = 1$ for all $i \in \mtcv$, the FJ model~\eqref{eq_M2} degenerates to the DeGroot model~\eqref{eq_M1}. 

The repelling negative dynamics (\cite{shi2019dynamics}) can capture the opinion evolution  where the network contains negative edges (i.e., $\mtce^- \not= \emptyset$): 
\begin{align}\nonumber
    &x_i(t+1) \\ \nonumber
    &= x_i(t) + \alpha_i \sum_{j \in \mtcn_i^+} (x_j(t) - x_i(t)) - \beta_i \sum_{j \in \mtcn_i^-} (x_j(t) - x_i(t)) \\\nonumber
    &= (1 - \alpha_i d_i^+ + \beta_i d_i^-) x_i(t) + \alpha_i \sum_{j \in \mtcn_i^+} x_j(t) - \beta_i \sum_{j \in \mtcn_i^-} x_j(t) \\ \label{eq_M3}
    &=: f^{\ModelRepell}(x(t),\theta^{(\ModelRepell,i)})
\end{align}
where $0<\alpha_i\le 1/d_i^+$ and $\beta_i > 0$ are the influence strength of positive and negative neighbors on the agent~$i$, respectively. Here we denote the parameters 
\begin{equation}\label{eq_theta3}
    \theta^{(\ModelRepell,i)} \Let [w_{i1}~\cdots~w_{in}]^\tp,~i\in\mtcv,
\end{equation} 
where $w_{ii} = 1 - \alpha_i d_i^+ + \beta_i d_i^-$, $w_{ij} = \alpha_i$ if $j \in \mtcn_i^+\setminus\{i\}$, $w_{ij} = -\beta_i$ if $j \in \mtcn_i^-$, and $w_{ij} = 0 $ if $j \not \in \mtcn_i$.  When $\mtcn_i^- = \emptyset$ or $\beta_i = 0$  for all $i\in \mtcv$, the model~\eqref{eq_M3} becomes the DeGroot model~\eqref{eq_M1}. 

Now we assume $\mtce^- = \emptyset$ again, and introduce the social HK model (\cite{
parasnis2018hegselmann}). In the original HK model
, the network is assumed to be complete, which is not realistic because individuals in a large network cannot know everyone. The social HK model addresses this issue by introducing an underlying network: At each time $t$, an agent~$i$ selects a set of trusted individuals from its neighbors (i.e., agents that have opinions similar to~$i$),
\begin{equation*}
    \mtci_i(t) = \{j \in \mtcn_i: |x_j(t) - x_i(t)| \le c_i \},
\end{equation*}
where $c_i$ is the confidence bound of~$i$. Then the agent updates its opinions as the average of  $x_j(t)$:
\begin{equation}\label{eq_M4}
    x_i(t+1) = \frac{1}{|\mtci_i(t)|} \sum_{j\in \mtci_i(t)} x_j(t) =: f^{\ModelHK}(x(t),\theta^{(\ModelHK,i)}), 
\end{equation}where 
\begin{equation}\label{eq_theta4}
    \theta^{(\ModelHK,i)} \Let c_i,~i\in\mtcv.
\end{equation}
Note that $f^{\ModelHK}$ depends on $i$, which is omitted for notation simplicity.
When all $c_i$ are larger than the range of $x(0)$, the model~\eqref{eq_M4} reduces to the DeGroot model~\eqref{eq_M1}. 

\section{Problem Formulation}\label{sec_prob}
Agents in real networks may have completely different  update rules, as suggested by empirical evidence (\cite{clemm2017micro,friedkin1999social,
kozitsin2021opinion,
kozitsin2022formal}). Thus, to learn the real network and the dynamics, it is natural to consider a mixture of opinion update rules~\cite{dong2017dynamics,wu2022mixed}.
A mixed opinion dynamics is defined as follows.
\begin{defn}\label{def}
    A mixed opinion dynamics over an undirected graph $\mtcg = (\mtcv,\mtce,A)$ with $|\mtcv|=n$ is a discrete-time system with states $x(t) \in \RR^n$, $t\in \NN$, satisfying that
    \begin{equation*}
        x_i(t+1) = f^{\ModelFori}(x(t),\theta^{(\ModelFori,i)}),
    \end{equation*}
    where $\ModelFori \in \{\ModelDeGroot, \ModelFJ, \ModelRepell, \ModelHK\}$ is the type of update rule of the agent~$i$, $f^{\ModelFori}$ are given in~\eqref{eq_M1},~\eqref{eq_M2},~\eqref{eq_M3}, and~\eqref{eq_M4}, and $\theta^{(\ModelFori,i)}$    in~\eqref{eq_theta1},~\eqref{eq_theta2},~\eqref{eq_theta3}, and~\eqref{eq_theta4}.
\end{defn}

To make the definition well-posed, we introduce the following assumptions.
\begin{assum}\label{asmp}   ~\\
    (i) The graph $\mtcg$ is undirected and connected.\\
    (ii) There exists a positive constant $\eps_\lambda \in (0,1)$ such that, if $\ModelFori = \ModelFJ$ for some $i\in\mtcv$, then $\eps_\lambda \le \theta^{(\ModelFJ,i)}_{n+1} \le 1 - \eps_\lambda$.\\
    (iii) For $i\in \mtcv$,  $\mtcn_i^- \not= \emptyset$ holds if and only if $\ModelFori = \ModelRepell$.\\ 
    (iv) If $\ModelFori = \ModelHK$ for $i\in\mtcv$, then $\theta^{(\ModelHK,i)} < \max_{j\in\mtcv}\{|x_j(0)|\}$.
\end{assum}

\begin{rem}
    The second assumption 
    ensures that the susceptibility of an agent~$i$ 
    should be neither too small nor too large. 
    The third assumption guarantees that 
    agents who have other update rules do not have negative edges
    . 
    The last assumption ensures that the agent with a HK update rule cannot be considered as a DeGroot-type agent.
\end{rem}

We consider the joint learning of the network topology and the dynamics for the mixed model:

\textbf{Problem.} Given a trajectory $\{x(0), x(1), \dots, x(T)\}$ of the mixed model, propose an algorithm jointly learning the network $\mtcg$, the types of agent update rules $\{\ModelFori\}$, and the parameters of each agent $\{\theta^{(\ModelFori,i)}\}$, where $T \ge 1$ is the final time step of the trajectory.


\section{Learning Algorithms}\label{sec_alg}

This section presents separate learning algorithms for models $\SetAllModels$, and proposes a learning algorithm for the mixed model. Denote the data matrices by $\bm{X}\Let [x(0)~x(1)~\dots~x(T-1)]^\tp$ and $b^{(i)} \Let [x_i(1)~\dots~x_i(T)]^\tp$, $ i\in \mtcv$, and the estimates of the adjacency matrix $A$, agent update rules $\{\ModelFori\}$, and the parameter $\{\theta^{(\ModelFori,i)}\}$ by $\hat{A}$, $\{\hat{\ModelFori}\}$, and $\{\hat{\theta}^{(\ModelFori,i)}\}$, respectively.

\begin{algorithm}[t]
\caption{Learn$\ModelDeGroot$($\bm{X}$,$b^{(i)}$,$\mtcn_i^{(\tx{neigh})}$,$\mtcn_i^{(\tx{non})}$)}\label{alg_DG}
{\fontsize{8}{8}\selectfont
\begin{algorithmic}[1]
\STATE {  For $y\in \RR^n$, solve
\begin{align*}
    \min ~&\|y\|_1\\
    \tx{s.t.} ~&\bm{X} y = b^{(i)},\\
    & \bfl^T y = 1,\\
    & y_j \ge \eps_{w}, ~j \in \mtcn_i^{(\tx{neigh})},\\
    & y_j = 0, ~j \in \mtcn_i^{(\tx{non})},\\
    & y_j \ge 0, ~j \in \mtcv \setminus (\mtcn_i^{(\tx{neigh})} \cup \mtcn_i^{(\tx{non})}).
\end{align*}}
\STATE {  Return $[\hat{A}]_{i,:} = \mathbb{I}_{[y^\tp > 0]}$, $\hat{\theta}^{(\ModelDeGroot,i)} = y$.}
\end{algorithmic}}
\end{algorithm}

\begin{algorithm}[t]
\caption{Learn$\ModelFJ$($\bm{X}$,$b^{(i)}$,$\mtcn_i^{(\tx{neigh})}$,$\mtcn_i^{(\tx{non})}$)}\label{alg_FJ}
{\fontsize{8}{8}\selectfont
\begin{algorithmic}[1]
\STATE {  For $y\in \RR^{n+1}$, solve
\begin{align*}
    \min ~&\|y\|_1\\
    \tx{s.t.} ~&[\bm{X}~[\bm{X}]_{1,i}\bfl_{T}] y = b^{(i)},\\
    & \bfl^T y = 1,\\
    & y_j \ge \eps_{w} \eps_{\lambda},~ j \in \mtcn_i^{(\tx{neigh})},\\
    & y_j = 0, ~j \in \mtcn_i^{(\tx{non})},\\
    & y_j \ge 0, ~j \in \mtcv \setminus (\mtcn_i^{(\tx{neigh})} \cup \mtcn_i^{(\tx{non})}),\\
    & \eps_{\lambda}\le y_{n+1} \le 1 - \eps_{\lambda}.
\end{align*}}
\STATE {  Return $[\hat{A}]_{i,:} = \mathbb{I}_{[y_{1:n}^\tp > 0]}$, $\hat{\theta}^{(\ModelFJ,i)}_{1:n} = y_{1:n}/(1-y_{n+1})$, $\hat{\theta}^{(\ModelFJ,i)}_{n+1} = 1- y_{n+1}$.}
\end{algorithmic}}
\end{algorithm}

\begin{algorithm}[t]
\caption{Learn$\ModelRepell$($\bm{X}$,$b^{(i)}$,$\mtcn_i^{(\tx{neigh})}$,$\mtcn_i^{(\tx{non})}$)}\label{alg_sign}
{\fontsize{8}{8}\selectfont
\begin{algorithmic}[1]
\STATE {  For $y\in \RR^n$, solve
\begin{align*}
    \min ~&\|y\|_1\\
    \tx{s.t.} ~&\bm{X} y = b^{(i)},\\
    & y_j \ge \eps_{w}, j \in \mtcn_i^{(\tx{neigh})},\\
    & y_j = 0, j \in \mtcn_i^{(\tx{non})}.
\end{align*}}
\STATE {  Return $[\hat{A}]_{i,:} = \sgn(y^\tp)$, $\hat{\theta}^{(\ModelRepell,i)} = y$.}
\end{algorithmic}}
\end{algorithm}

\subsection{Learning of Single Models}
Note that $\theta^{(\ModelDeGroot,i)}$ satisfies 
$\bm{X} \theta^{(\ModelDeGroot,i)} = b^{(i)}$, for $i\in\mtcv$ such that $\ModelFori = \ModelDeGroot$.
To learn $\theta^{(\ModelDeGroot,i)}$ we look for a sparse solution to the least $l_1$-norm problem given in Algorithm~\ref{alg_DG}. The motivation is that in practice only a few samples of the process are available ($T < n$) but real networks are often sparse (\cite{
ravazzi2017learning}).
The parameter constraints $\bfl^T y = 1$ and $y_j \ge 0$, $1\le j \le n$, are added.
The algorithm has two additional inputs $\mtcn_i^{(\tx{neigh})}$ and $\mtcn_i^{(\tx{non})}$, which are subsets of $\mtcv$. We introduce these two sets  to search for other solutions. If an agent~$j$ is considered as a neighbor of the agent~$i$ ($j \in \mtcn_i^{(\tx{neigh})}$), we add the constraint $y_j \ge \eps_{w}$, where $\eps_{w}$ is a small   positive constant
. If $j$ is assumed to not be a neighbor ($j \in \mtcn_i^{(\tx{non})}$), the constraint $y_j = 0$ is imposed.

We can similarly solve a system of linear equations to learn the parameters in the FJ model~\eqref{eq_M2}. 
We search for a sparse solution in Algorithm~\ref{alg_FJ}. The algorithm has two hyperparameters: the lower bound of influence weights $\eps_{w}>0$, and the bound for the susceptibility $\eps_{\lambda} \in (0,1)$ given in Assumption~\ref{asmp}~(ii).

The expression of the repelling negative dynamics~\eqref{eq_M3} is the same as~\eqref{eq_M1}, but~\eqref{eq_M3} has less constraints. Hence so does Algorithm~\ref{alg_sign}. 
In Section~\ref{sec_improv} we introduce more constraints for this algorithm.

\begin{algorithm}[t]
\caption{Learn$\ModelHK$($\bm{X}$,$b^{(i)}$,$\mtcn_i^{(\tx{neigh})}$,$\mtcn_i^{(\tx{non})}$)}\label{alg_HK}
{\fontsize{8}{8}\selectfont
\begin{algorithmic}[1]
\STATE {Set $[\hat{A}]_{i,:}$ such that $[\hat{A}]_{i,j} = 1$ if $j \in \mtcn_i^{(\tx{neigh})}$ and $[\hat{A}]_{i,j} = 0$ otherwise.}
\FOR{ $t$ from $0$ to $T-1$}
    \STATE {Sort the neighbors~$j \in \mtcn_i^{(\tx{neigh})}$ by the distance of $x_j(t)$ from $x_i(t)$ to be 
    $x_{i_1}$, $\dots$, $x_{i_{n_i}}$, where $n_i \Let [\hat{A}]_{i,:}\bfl$.}
    \STATE {Compute
    \begin{align*}
        \hat{x}_i^{(m)}(t+1) &= \frac{1}{m} \sum_{j=1}^m x_{i_j}(t),~1\le m \le n_i,\\
        m(t) &= \max \bigg\{ \underset{1\le m \le n_i}{\arg\min} | \hat{x}_i^{(m)}(t+1) - x_i(t+1)  | \bigg\},\\
        c(t) &= |x_{i_{m(t)+1}}(t) - x_i(t)| - \eps_c.
    \end{align*}}
\ENDFOR
    \STATE {~Compute
    \begin{align*}
        t^* &= \underset{0\le t\le T-1}{\arg\min} c(t),\\
        c^* &= c(t^*).
    \end{align*}}
\STATE { Return $[\hat{A}]_{i,:}$, $\hat{\theta}^{(\ModelHK,i)} = c^*$.
}
\end{algorithmic}}
\end{algorithm}

\begin{algorithm}[t]
\caption{$\eps$-Greedy($\{x(0),\dots,x(T)\}$)}\label{alg_eps}
{\fontsize{8}{8}\selectfont
\begin{algorithmic}[1]
\STATE Choose $\tilde{T}$ and divide the trajectory into a training dataset $\bm{X}_{\tx{tr}} = [x(0)~\cdots~x(\tilde{T})]^\tp$ and a validation dataset $\bm{X}_{\tx{val}} = [x(\tilde{T}+1)~\cdots~x(T)]^\tp$.
\STATE Initialize the $Q$-table 
$Q(0)$, the estimates of update rule types $\hat{k}(0)$, the estimates of the adjacency matrix for each model $\hat{A}^{(M[m])}(0)$, and the parameter estimates $\hat{\theta}^{(M[m],i)}(0)$, $1\le m\le 4$, where the look-up table $M$ is s.t. $M[1]=\ModelDeGroot$, $M[2]=\ModelFJ$, $M[3]=\ModelRepell$, $M[4]=\ModelHK$. 
\FOR{$l$ from $1$ to $n_{\tx{iter}}$}
    \FOR{$i$ from $1$ to $n$}
        \STATE With probability (w.p.) $1-\eps_{\tx{M}}$, set
        \begin{equation*}
            k_i(l) = \underset{1\le m \le 4}{\arg\max} [Q(l)]_{i,m}.
        \end{equation*}
        \STATE W.p. $\eps_{\tx{M}}$, generate $k_i(l) \in\{1,2,3,4\}$ uniformly. 
    \ENDFOR
    \STATE W.p. $1 - \eps_{\tx{G}}$, let $[\tilde{A}]_{i,:} \Let [\hat{A}^{(k_i(l))}(l-1)]_{i,:}$, for $i\in\mtcv$. W.p. $\eps_{\tx{G}}$, randomly generate an adjacency matrix $\tilde{A}$.
    \FOR{$i$ from $1$ to $n$}
        \FOR{$j$ from $1$ to $n$ and $j\not= i$}
            \STATE If $\tilde{A}_{ij} = 0$ set $\mtcn_i^{\tx{(neigh)}} \Let \{q \in \mtcv: \tilde{A}_{iq} = 1\} \cup \{j\}$ and $\mtcn_i^{\tx{(non)}} \Let \emptyset$. If $\tilde{A}_{ij} = 1$ set $\mtcn_i^{\tx{(neigh)}} \Let \{q \in \mtcv: \tilde{A}_{iq} = 1\} \setminus \{j\}$ and $\mtcn_i^{\tx{(non)}} \Let \{j\}$. 
            \begin{align*}&([\hat{A}_{\tx{temp}_j}]_{i,:},\hat{\theta}_{\tx{temp}_j}^{(M[k_i(l)],i)}) \\
            =& \tx{Learn}M[k_i(l)]([\bm{X}_{\tx{tr}}]_{1:\tilde{T}-1,:},[\bm{X}_{\tx{tr}}]_{2:\tilde{T},i}, \mtcn_i^{\tx{(neigh)}},\mtcn_i^{\tx{(non)}})
            \end{align*}
        \ENDFOR
        \STATE Compute the validation error of $f^{M[k_i(l)]}$ based on $([\hat{A}_{\tx{temp}_j}]_{i,:},\hat{\theta}_{\tx{temp}_j}^{(M[k_i(l)],i)})$, $j \in \mtcv\setminus \{i\}$. Choose the smallest error $p_i(l)$ with respect to $j$, update $[\hat{A}^{(M[k_i(l)])}(l)]_{i,:}$, $\hat{\theta}^{(M[k_i(l)],i)}(l)$ accordingly, $[\hat{A}^{(M[m])}(l)]_{i,:} = [\hat{A}^{(M[m])}(l-1)]_{i,:}$ and $\hat{\theta}^{(M[m],i)}(l) = \hat{\theta}^{(M[m],i)}(l-1)$ for $m = \{1,2,3,4\}\setminus\{k_i(l)\}$, and
        \[[Q(l)]_{i,k_i(l)} = (1 - \alpha)[Q(l-1)]_{i,k_i(l)} - \alpha\log(p_i(l)).\] 
    \ENDFOR
\ENDFOR
\STATE Let $m_i \Let \underset{1\le m\le 4}{\arg\max} [Q(n_{\tx{iter}})]_{i,m}$, $\hat{\ModelFori} \Let M[m_i]$, $[\hat{A}]_{i,:} = [\hat{A}^{(\hat{\ModelFori})} (n_{\tx{iter}})]_{i,:}$, and $\hat{\theta}^{(\hat{\ModelFori},i)} = \hat{\theta}^{(\hat{\ModelFori},i)}(n_{\tx{iter}})$, $i\in \mtcv$.
\STATE Return $\hat{A}$, $\{\hat{\ModelFori}\}$, and $\{\hat{\theta}^{(\hat{\ModelFori},i)}\}$.
\end{algorithmic}
}
\end{algorithm}

For the Social HK model, it is only possible to obtain a bound for the confidence bounds based on finite samples. We provide a heuristic to estimate the confidence bound. Numerical experiments in Section~\ref{sec_simulation} show that the proposed algorithm can have relatively small prediction error. 
Assume first that the network topology is known. For an agent~$i$ with $\ModelFori = \ModelHK$, sort~$i$'s neighbors by their states' distance from $x_i(t)$ and denote the sorted neighbors by $i_1,i_2, \dots,i_{|\mtcn_i|}$, where $i_1 = i$. Then compute the averages $(\sum_{j=1}^m x_{i_j}(t))/m$, $1\le m \le |\mtcn_i|$, and find the index such that the average has the smallest distance from $x_i(t$$+1)$, i.e., ${\arg\min}_{1\le m\le |\mtcn_i|} |( \sum_{j=1}^m x_{i_j}(t))/m - x_i(t+1) |$.
If no noise exists, there exist $m_1< \dots <m_q$ such that $|(\sum_{j=1}^{m_r} x_{i_j}(t))/m_r - x_i(t+1)| = 0$, where $1\le r\le q$ and $q \ge 1$. We then can conclude that $|x_{i_{m_{q+1}}}(t) - x_i(t)|$ is a strict upper bound for the confidence bound $\theta^{(\ModelHK, i)} = c_i$. 
If the topology is unknown, to obtain a bound we need test all $2^{n-1}$ combinations
. We narrow down the search by introducing testing sets $\mtcn_i^{(\tx{neigh})}$ and $\mtcn_i^{(\tx{non})}$, as in the preceding algorithms. 
In Algorithm~\ref{alg_HK}, 
for each sample $(x(t),x_i(t+1))$, we obtain an estimate of the confidence bound $c(t)$ by conducting the process discussed previously. In the algorithm, $\eps_{c}$ is a small positive number to make sure that the upper bound is strict. For all~$0\le t\le T-1$, if the neighbor set is correct, the confidence bound should be smaller than all $c(t)$.  The algorithm returns an approximation of~$i$'s dynamics.

\subsection{Learning of Mixed Model}\label{subsec_epsG}
After presenting the learning algorithms for each update rule, now we are ready to introduce the learning algorithm for the mixed model. 
We propose a multi-armed bandit algorithm (Algorithm~\ref{alg_eps}) to address the problem, and consider the four types of update rules as four arms. We introduce a $Q$-table $Q(l) \in \RR^{n\times 4}$ for each iteration $l$. The entry $[Q(l)]_{i,m}$ indicates the payoff of refining model parameters of the arm $M[m]$ for the agent~$i$ at the iteration $l$, where $1\le m \le 4$ and we define the look-up table $M$ with respect to $\SetAllModels$ such that $M[1]=\ModelDeGroot$, $M[2]=\ModelFJ$, $M[3]=\ModelRepell$, $M[4]=\ModelHK$.
We run Algorithms~\ref{alg_DG}--\ref{alg_HK} for each~$i$ with $\mtcn_i^{(\tx{neigh})} = \{i\}$ and $\mtcn_i^{(\tx{non})} = \emptyset$, to initialize the estimates of update rule types $\hat{k}(0) \in \{1,2,3,4\}^n$, the estimates of the adjacency matrix for each model $\hat{A}^{(M[m])}(0)$, and the parameter estimates $\hat{\theta}^{(M[m],i)}(0)$. $[Q(0)]_{i,m}$ is set to be the negative logarithm of validation error of the model $M[m]$ for~$i$  (Line~$2$). 

At each iteration and for each agent, the algorithm tries to fit the data into the model with the best payoff with probability $1 - \eps_{\tx{M}}$. With exploration probability $\eps_{\tx{M}}$, the algorithm attempts to refine estimates of other models (Line~$4$--$7$). 
At each iteration, the algorithm modifies the adjacency matrix corresponding to the selected models with probability $1 - \eps_{\tx{G}}$, and modifies a randomly generated adjacency matrix with probability $\eps_{\tx{G}}$ (Line~$8$). 
The algorithm then examines whether removing or adding an edge can improve validation error (Line~$11$). Next, the algorithm updates the payoff $Q$ with a step-size $\alpha$ and the parameter estimates (Line~$13$). Finally the algorithm returns parameter estimates, according to the update rule with the best payoff (Line~$17$).


\begin{figure*}
\centering
    \subfigure[\label{roc10}$T=10$]{    \includegraphics[width=0.17\linewidth]{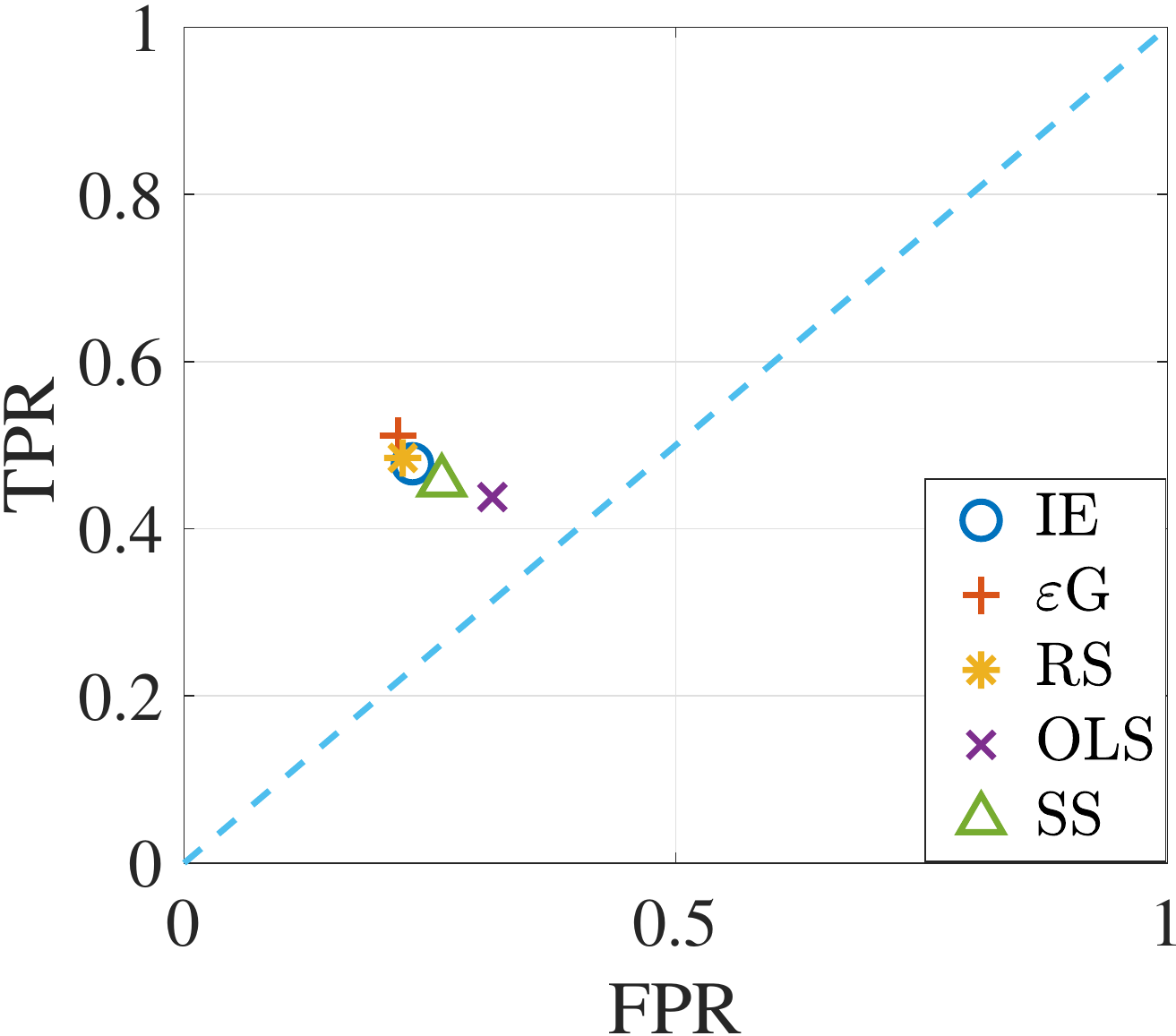}}\qquad
    \subfigure[\label{roc15}$T=15$]{\includegraphics[width=0.17\linewidth]{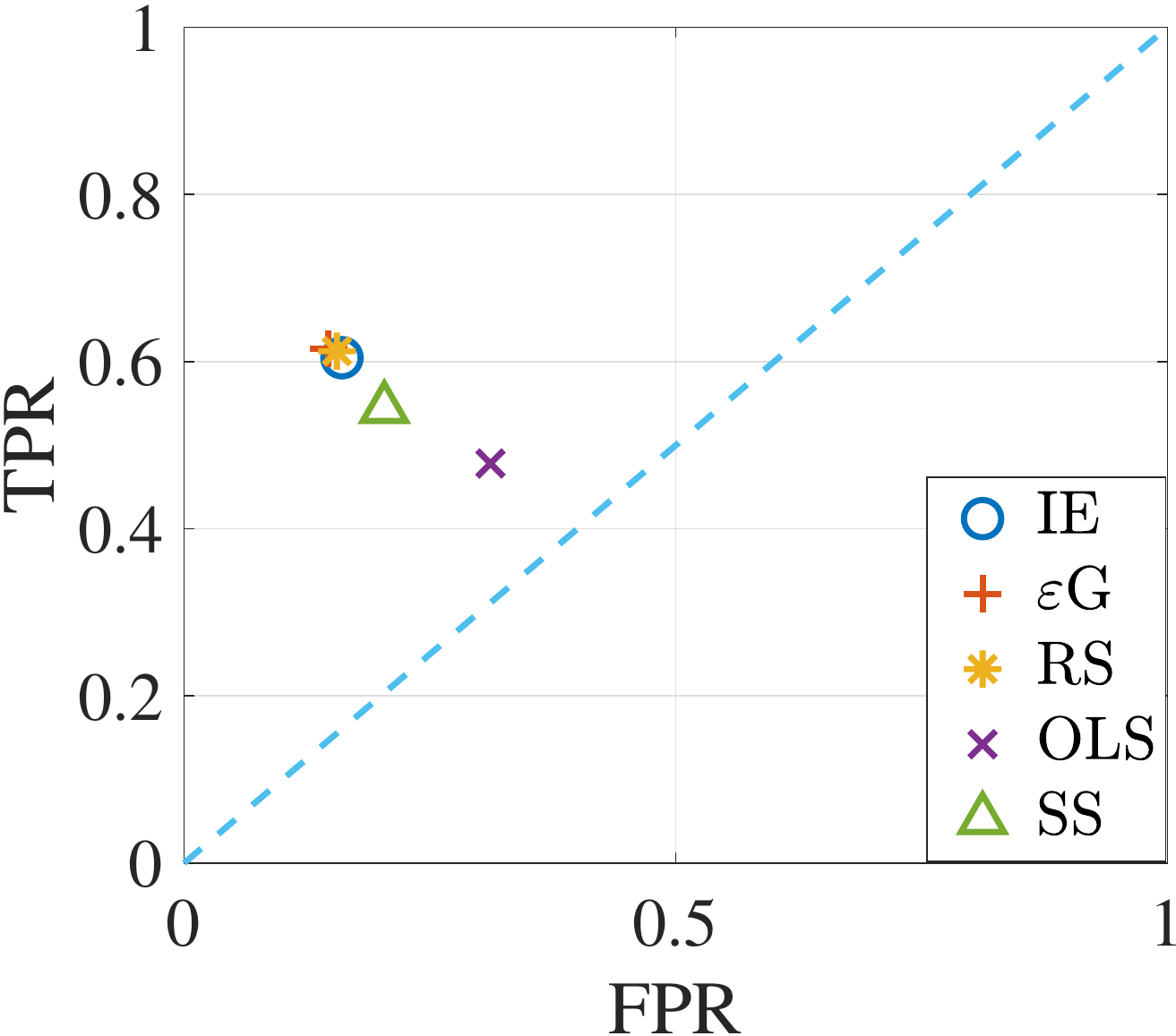} }\qquad
    \subfigure[\label{roc20}$T=20$]{\includegraphics[width=0.17\linewidth]{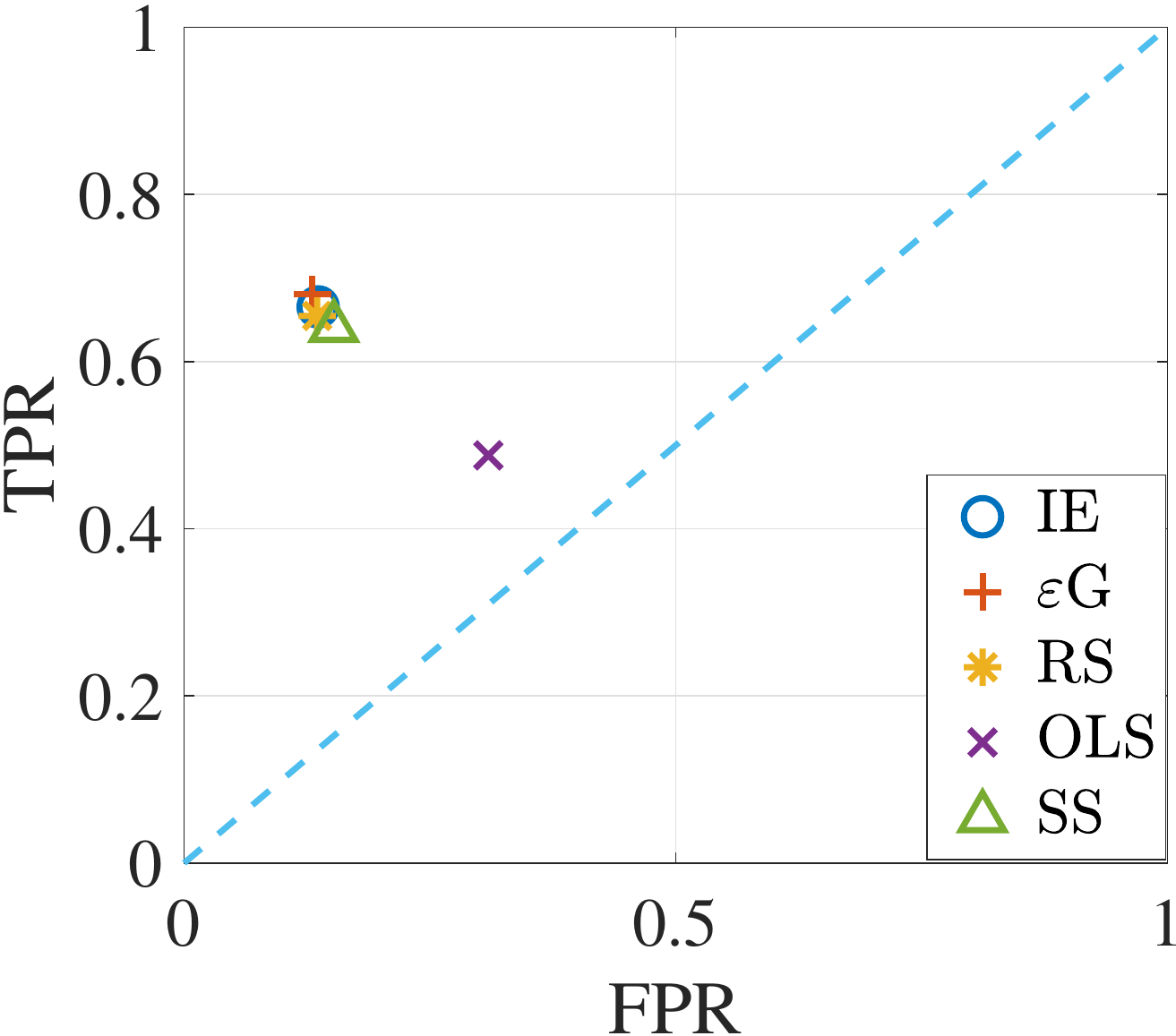} }
    \caption{\label{fig_roc} Performance of algorithms IE, $\eps$G, RS, OLS, and SS for network topology recovery.} 
\end{figure*}

\begin{figure*}
\centering
    \subfigure[\label{boxp10}$T=10$]{    \includegraphics[width=0.2\linewidth]{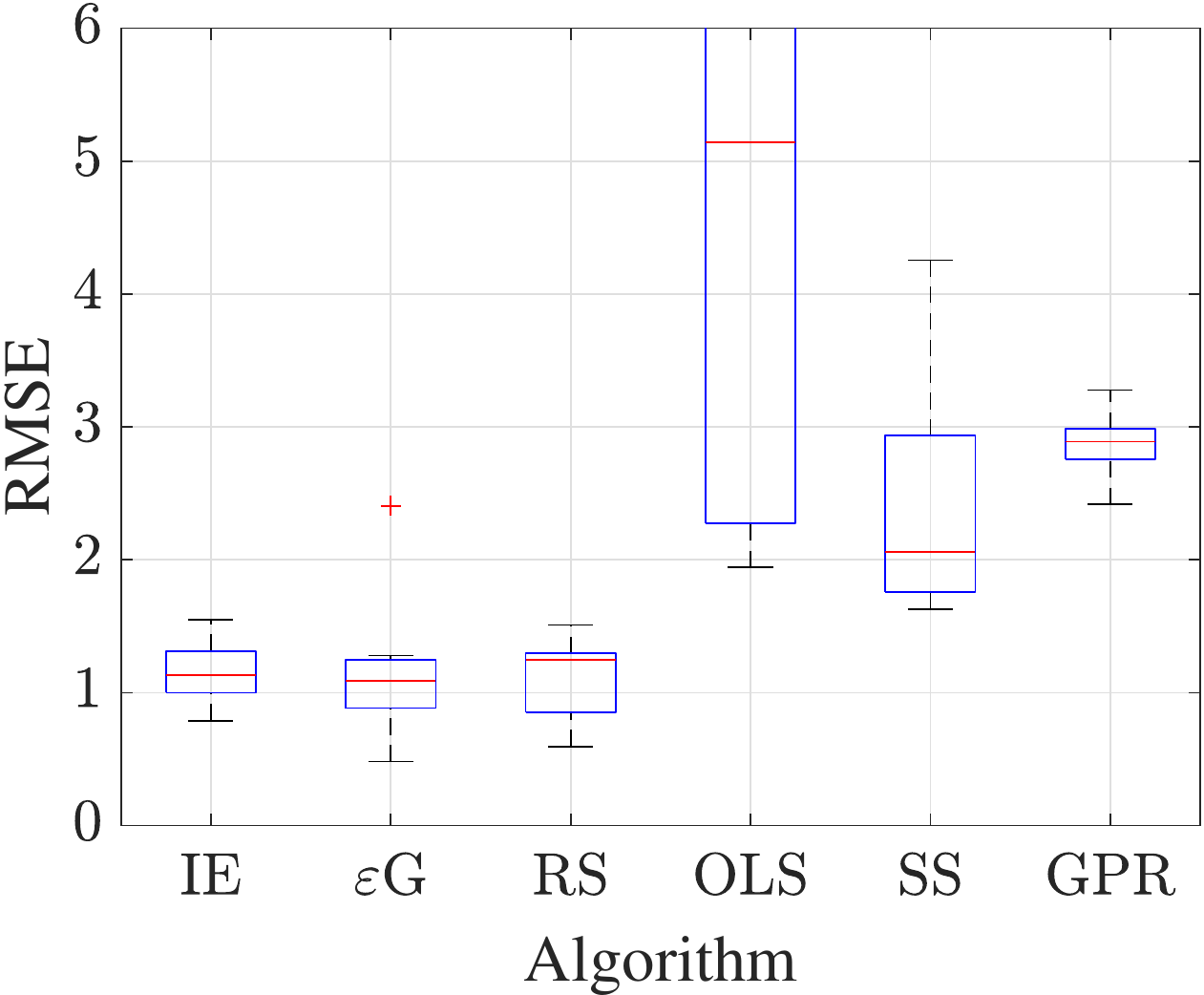}}\quad
    \subfigure[\label{boxp15}$T=15$]{\includegraphics[width=0.2\linewidth]{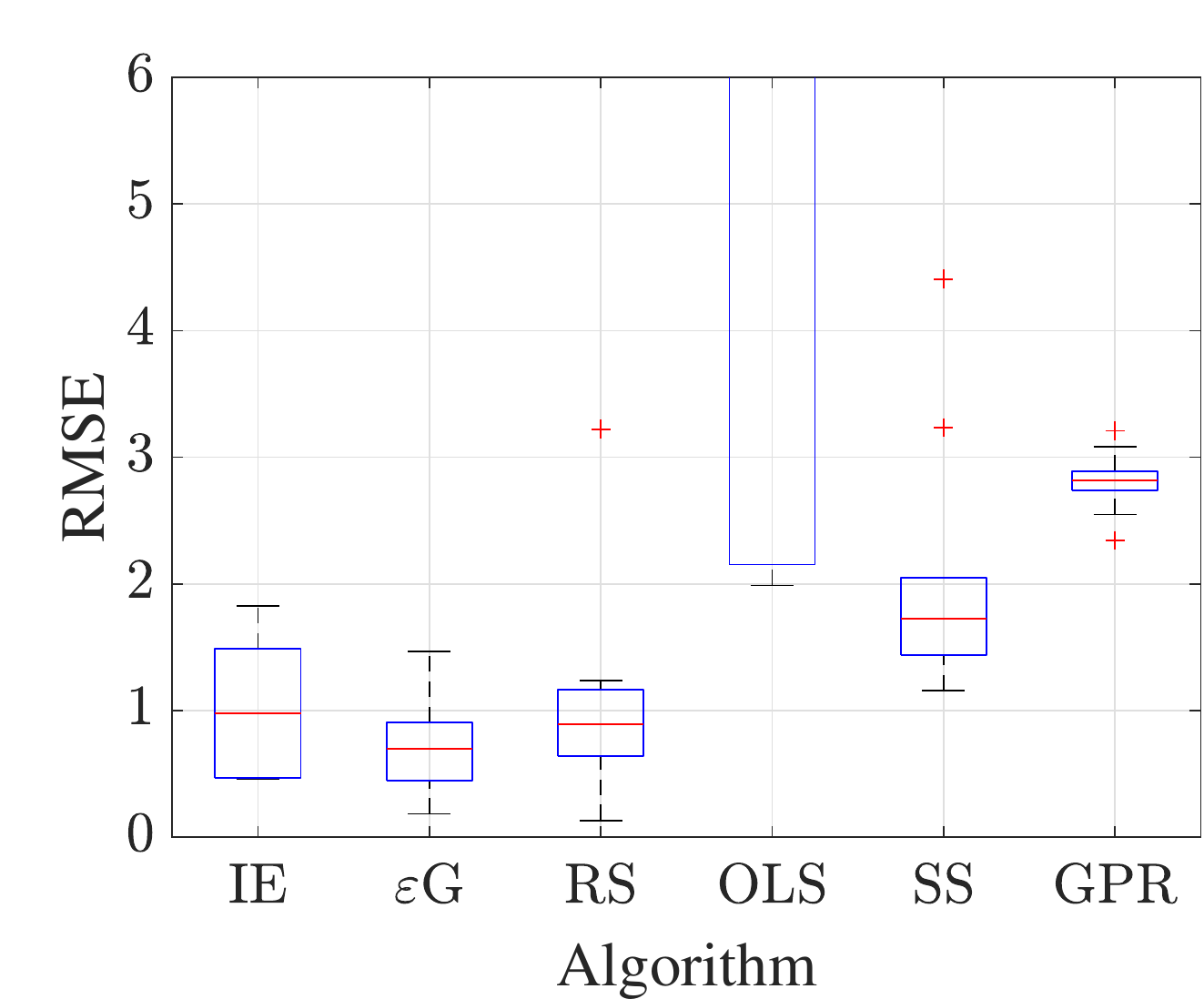} }\quad
    \subfigure[\label{boxp20}$T=20$]{\includegraphics[width=0.2\linewidth]{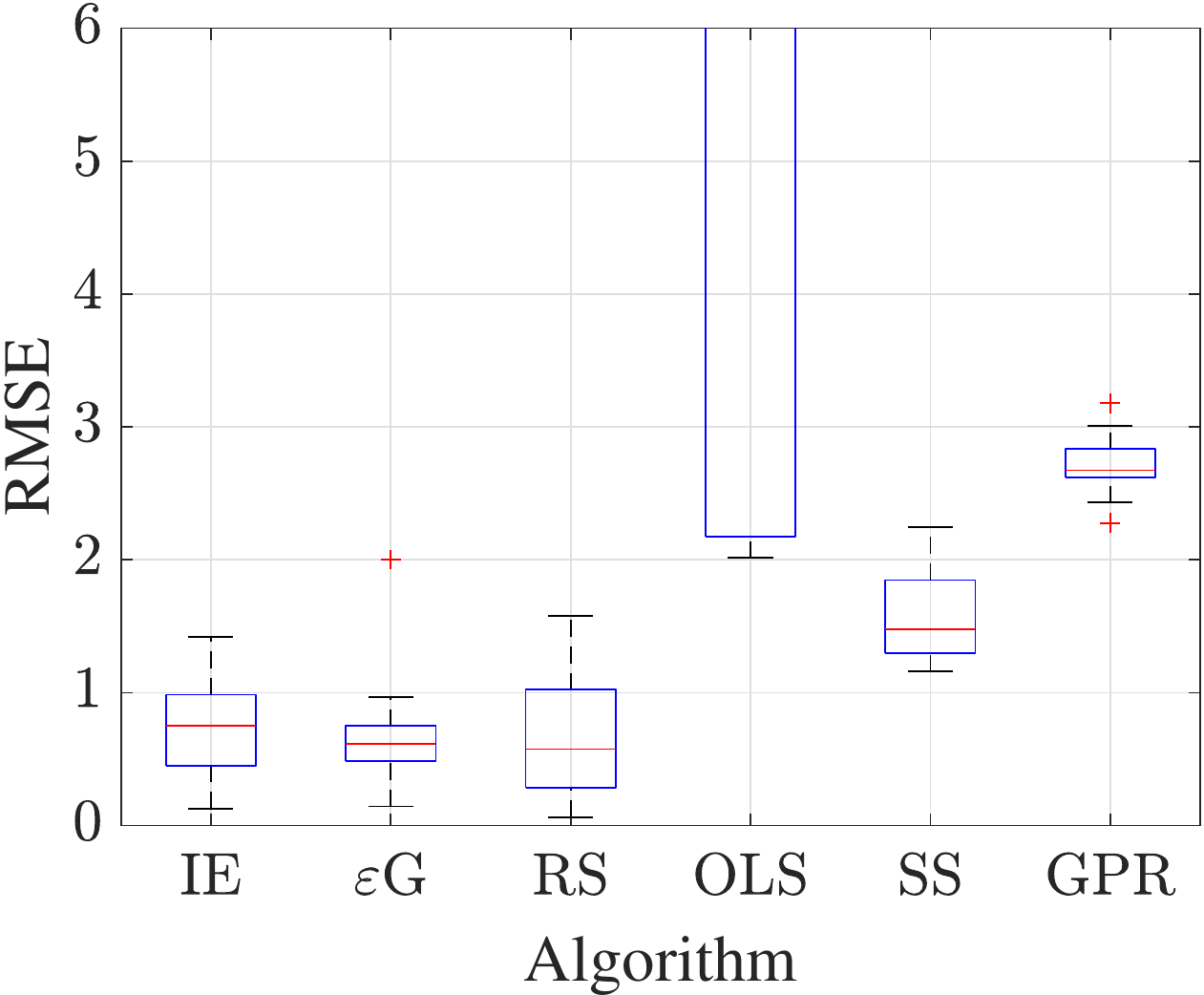} }
    \caption{\label{fig_bospl} Prediction error of algorithms IE, $\eps$G, RS, OLS, SS, and GPR.} 
\end{figure*}

\section{Numerical Experiments}\label{sec_simulation}

\subsection{Performance of Algorithm~\ref{alg_eps}}
This section presents a numerical experiment for illustration of algorithm performance. 
We use the CVX toolbox\footnote{\url{http://cvxr.com/cvx}} to find sparse solutions.

To generate the mixed model, we set the network size to be $n=20$, with four types of agents and each type with~$5$ agents sharing the same model as ground truth. We generate $10$ graphs and run the model over each graph to get a trajectory with a final time step $t=20$. The positive graphs are generated from an Erd{\"o}s-R{\'e}nyi graph with link probability $(1.1\log n)/n$ and are checked to be connected. Generate negative edges between the $5$ agents with the repelling rule, with the same link probability, and ensure that each agent has at least one negative edge. The influence weights $\alpha_i$ and $\beta_i$ are set to be $0.2/d_i$. The susceptibility $\lambda_i$ of the agents with the FJ rule is set to be $0.5$, and the confidence bound $c_i$ of the social HK model is set to be $0.25$. The initial opinions of the agents are generated independently and uniformly from $(-1,1)$.

We compare the proposed algorithm (Algorithm~\ref{alg_eps}, $\eps$G) with an initial estimate (IE) of the mixed model, random search (RS, Algorithm~\ref{alg_eps} with model and topology exploration probabilities $\eps_M = 1$ and $\eps_G = 1$), the ordinary least-squares algorithm (OLS), sparse solutions (SS) to $\bm{X}y = b$, and Gaussian process regression (GPR). For IE, we run Algorithm~\ref{alg_DG}--\ref{alg_HK} (with $\eps_w = 0.001$, $\eps_\lambda = 0.1$, and $\eps_c = 10^{-6}$) for each agent, and obtain initial estimates of update rules, model parameters, and the adjacency matrix. $\eps$G starts with the estimates of IE
, the number of iterations is set to be $20$, both $\eps_M$ and $\eps_G$ are set to be $0.2$, respectively, and the step size is $0.1$. 
For OLS, SS, and GPR, the data matrix for an agent~$i$ is $\begin{bmatrix}
        x(0) & \cdots & x(T-1)\\
        x_i(0) & \cdots & x_i(0) \end{bmatrix}^\tp$, thus including the effect of initial states in the FJ model.~Here $T$ is the number of samples.
For OLS and SS, we search solutions in $[-2,2]^{n+1}$ to avoid large weight estimates. To investigate the effect of sample number, for each $T = 10, 15, 20$, we apply the algorithms to the $10$ trajectories $\{x(0), \dots, x(T)\}$ (for $\eps$G, set $\tilde{T} = [4T/5]$), and study the averaged performance.

First, we examine the recovery of adjacency matrices by EI, $\eps$G, RS, OLS, and SS. The performance of the recovery is characterized by the the true positive rate (TPR) and the false positive rate (FPR) (see \cite{bishop2006pattern}). 
Fig.~\ref{fig_roc} shows the pair $(\tx{FPR},\tx{TPR})$ for the algorithms. As the number of samples increase, all algorithms perform better. $\eps$G is slightly better then IE when $T$ is small, indicating refinement of the initial estimates. SS has similar FPR and TPR to IE, which may be because SS also searches for sparse networks. But SS does not learn the model well and has large prediction error, as discussed later.

Next, we compute the prediction error for all algorithms. For each graph, we generate $50$ time-adjecent state pairs $\{x^{(g)}(0)),(x^{(g)}(1))\}$, $1\le g \le 50$. The prediction error is defined by the root mean square error (RMSE) $[(\sum_{g=1}^{50} \|\hat{x}^{(g)}(1) - x^{(g)}(1)\|^2)/50]^{\frac12}$,
where $\hat{x}^{(g)}(1)$ is the prediction of the states $x^{(g)}(1)$. The results are shown in Fig.~\ref{fig_bospl} with boxplots. IE, $\eps$G, and RS have smaller prediction error than the rest, because they are equipped with more model information. When the number of samples is small ($T=10$), $\eps$G and RS have better performance for some trajectories but they do not improve IE too much. Improvement can be observed in the case $T=15$, where $\eps$G has smaller error than IE and RS. When $T=20$, the performance of $\eps$G is similar to IE and RS, since there is sufficient information for all algorithms. OLS cannot capture the nonlinear dynamics so it performs the worst. SS and GPR has less prediction error than OLS, because SS finds sparse solutions, resulting in less error
, and GPR can capture nonlinear dynamics.

\begin{figure*}
\centering
    \subfigure[\label{mod10}$T=10$]{    \includegraphics[width=0.18\linewidth]{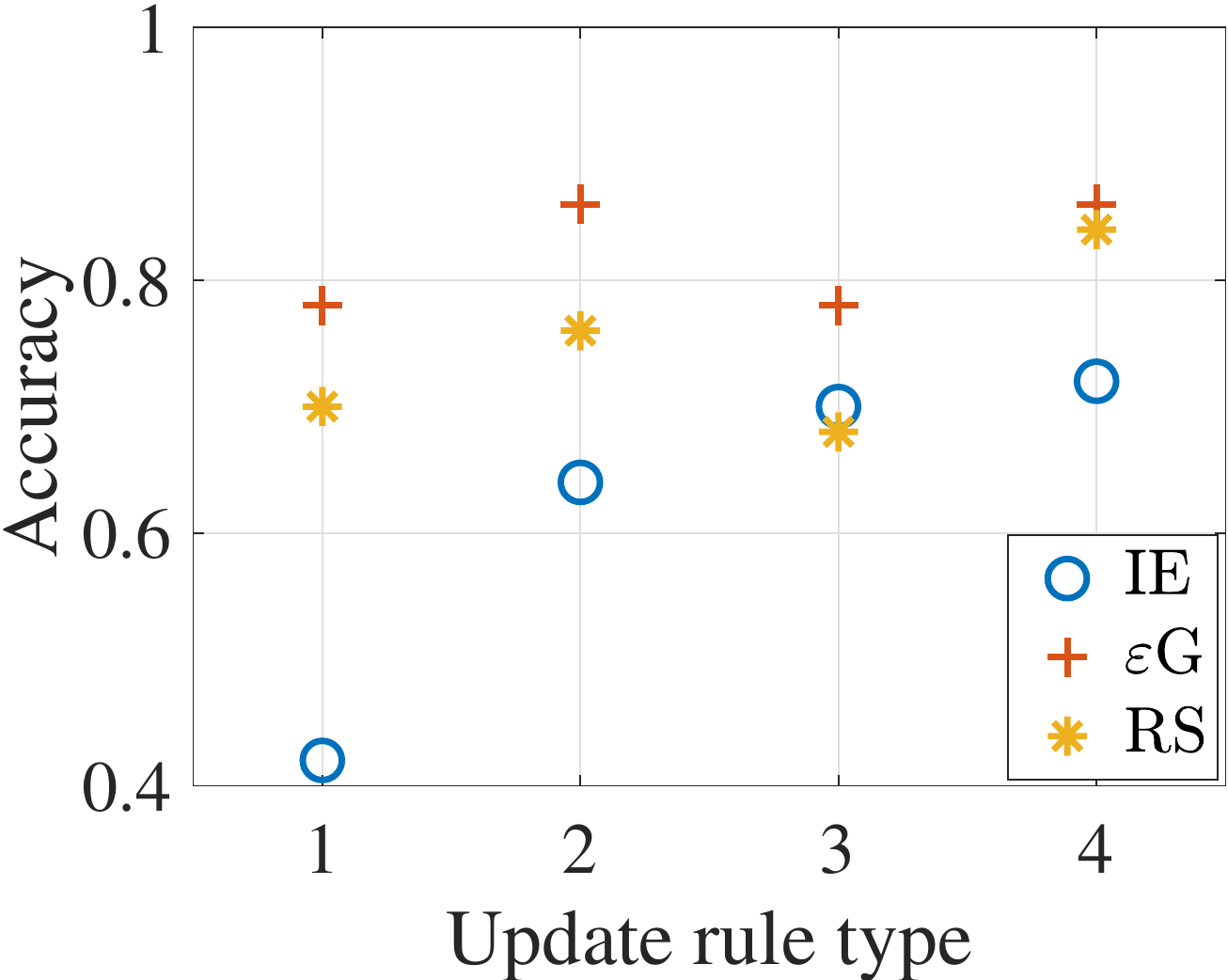}}\quad\quad
    \subfigure[\label{mod15}$T=15$]{\includegraphics[width=0.18\linewidth]{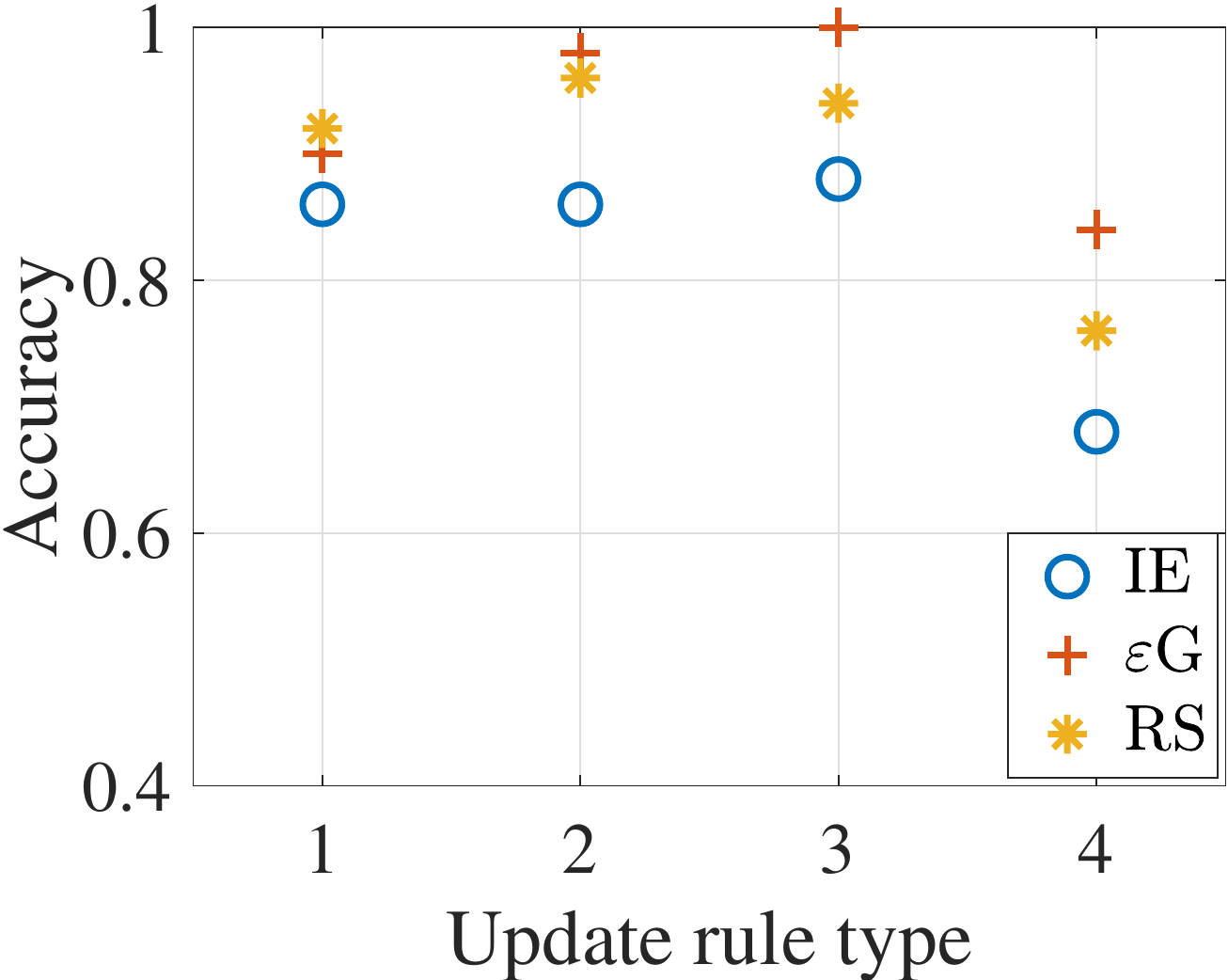} }\quad\quad
    \subfigure[\label{mod20}$T=20$]{\includegraphics[width=0.18\linewidth]{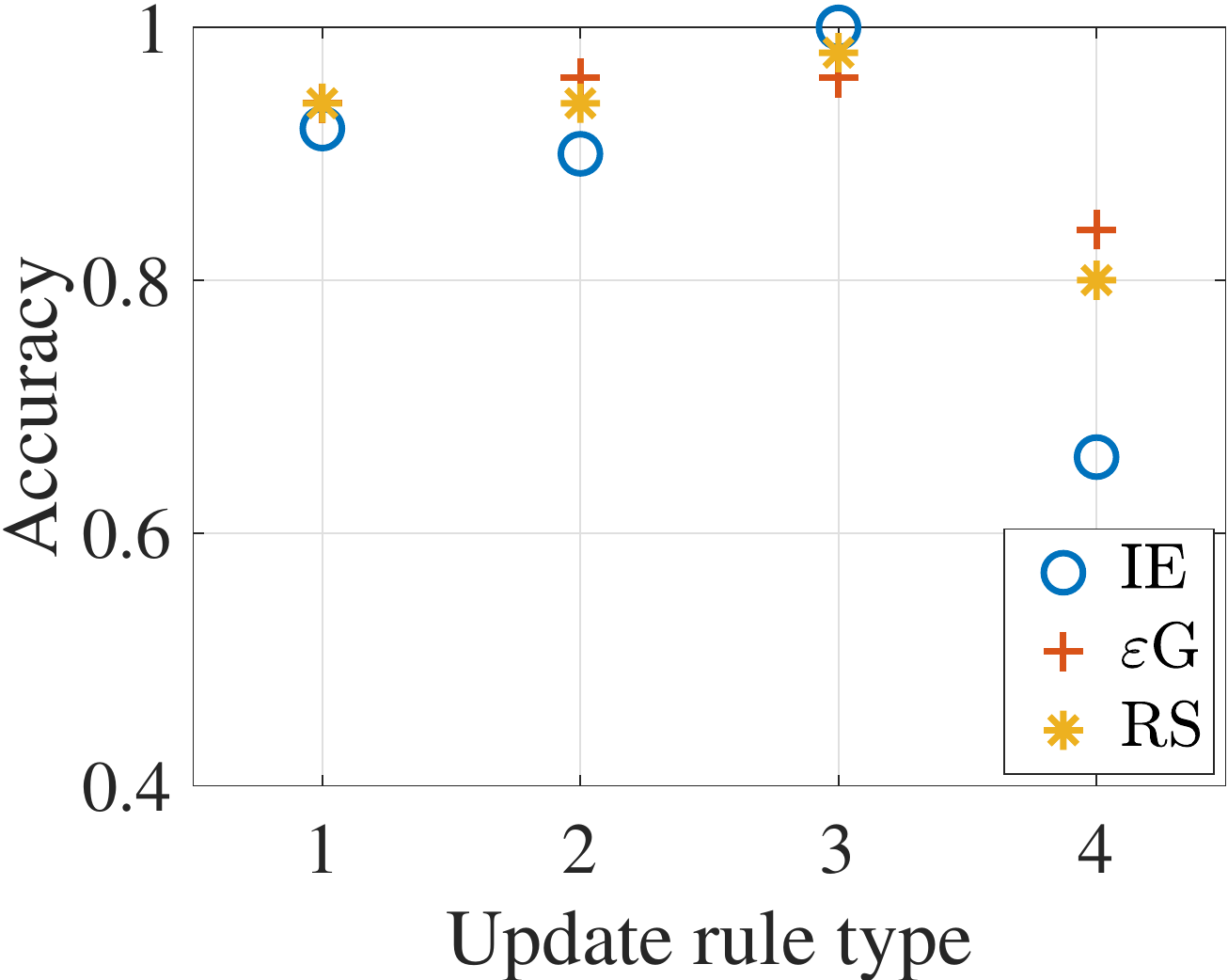} }
    \caption{\label{fig_mod} Accuracy of algorithms IE, $\eps$G, and RS for learning of update rule types.} 
\end{figure*}

We illustrate how the estimates of update rule types and parameters are influenced by the number of samples. Fig.~\ref{fig_mod} plots the accuracy (the proportion of correct estimates)  of IE, $\eps$G, and RS for each update rule type. The accuracy increases with the number of samples. $\eps$G has similar accuracy to RS, so the smaller prediction error may result from the exploitation for correctly learned update rules. 


\begin{algorithm}[t]
\caption{Learn$\ModelDeGroot^+$($\bm{X}$,$b^{(i)}$,$\mtcn_i^{(\tx{neigh})}$,$\mtcn_i^{(\tx{non})}$)}\label{alg_DG+}
{\fontsize{8}{8}\selectfont
\begin{algorithmic}[1]
\STATE {  For $y\in \RR^n$, solve
\begin{align*}
    \min_y ~&\|\bm{X} y - b^{(i)}\|^2\\
    \tx{s.t.} ~& \bfl^T y = 1,\\
    & y_j \ge \eps_{w}, ~j \in \mtcn_i^{(\tx{neigh})},\\
    & y_j = 0, ~j \in \mtcn_i^{(\tx{non})}.
\end{align*}}
\STATE {  Return $[\hat{A}]_{i,:} = \mathbb{I}_{[y^\tp > 0]}$, $\hat{\theta}^{(\ModelDeGroot,i)} = y$.}
\end{algorithmic}}
\end{algorithm}

\begin{algorithm}[t]
\caption{Learn$\ModelFJ^+$($\bm{X}$,$b^{(i)}$,$\mtcn_i^{(\tx{neigh})}$,$\mtcn_i^{(\tx{non})}$)}\label{alg_FJ+}
{\fontsize{8}{8}\selectfont
\begin{algorithmic}[1]
\STATE {  For $y\in \RR^{n+1}$, solve
\begin{align*}
    \min_y ~&\|[\bm{X}~[\bm{X}]_{1,i}\bfl_{T}] y - b^{(i)}\|^2\\
    \tx{s.t.} ~& \bfl^T y = 1,\\
    & y_j \ge \eps_{w} \eps_{\lambda},~ j \in \mtcn_i^{(\tx{neigh})},\\
    & y_j = 0, ~j \in \mtcn_i^{(\tx{non})},\\
    & \eps_{\lambda}\le y_{n+1} \le 1 - \eps_{\lambda}.
\end{align*}}
\STATE {  Return $[\hat{A}]_{i,:} = \mathbb{I}_{[y_{1:n}^\tp > 0]}$, $\hat{\theta}^{(\ModelFJ,i)}_{1:n} = y_{1:n}/(1-y_{n+1})$, $\hat{\theta}^{(\ModelFJ,i)}_{n+1} = 1- y_{n+1}$.}
\end{algorithmic}}
\end{algorithm}

\begin{algorithm}[t]
\caption{Learn$\ModelRepell^+$($\bm{X}$,$b^{(i)}$,$\mtcn_i^{(\tx{neigh})}$,$\mtcn_i^{(\tx{non})}$)}\label{alg_sign+}
{\fontsize{8}{8}\selectfont
\begin{algorithmic}[1]
\STATE {  For $y\in \RR^n$, solve
\begin{align*}
    \min_y ~&\|\bm{X} y - b^{(i)}\|^2\\
    \tx{s.t.} ~& \bfl^T y = 1,\\
    ~& y_j = 0, ~j \in \mtcn_i^{(\tx{non})}.
\end{align*}}
\STATE {  Return $[\hat{A}]_{i,:} = \sgn(y^\tp)$, $\hat{\theta}^{(\ModelRepell,i)} = y$.}
\end{algorithmic}}
\end{algorithm}

\begin{algorithm}[t]
\caption{$\eps$-Greedy$^+$($\{x(0),\dots,x(T)\}$)}\label{alg_eps+}
{\fontsize{8}{8}\selectfont
\begin{algorithmic}[1]
\STATE Choose $\tilde{T}$ and divide the trajectory into a training dataset $\bm{X}_{\tx{tr}} = [x(0)~\cdots~x(\tilde{T})]^\tp$ and a validation dataset $\bm{X}_{\tx{val}} = [x(\tilde{T}+1)~\cdots~x(T)]^\tp$.
\STATE Initialize the $Q$-table 
$Q(0)$, the estimates of update rule types $\hat{k}(0)$, the estimates of the adjacency matrix for each model $\hat{A}^{(M[m])}(0)$, and the parameter estimates $\hat{\theta}^{(M[m],i)}(0)$, $1\le m\le 4$, where $M[1]=\ModelDeGroot$, $M[2]=\ModelFJ$, $M[3]=\ModelRepell$, $M[4]=\ModelHK$. 
\FOR{$l$ from $1$ to $n_{\tx{iter}}$}
    \FOR{$i$ from $1$ to $n$}
        \STATE W.p. $1-\eps_{\tx{M}}$, set
        \begin{equation*}
            k_i(l) = \underset{1\le m \le 4}{\arg\max} [Q(l)]_{i,m}.
        \end{equation*}
        \STATE W.p. $\eps_{\tx{M}}$, generate $k_i(l) \in\{1,2,3,4\}$ uniformly. 
    \ENDFOR
    \STATE W.p. $1 - \eps_{\tx{G}}$, let $[\tilde{A}]_{i,:} \Let [\hat{A}^{(k_i(l))}(l-1)]_{i,:}$, for $i\in\mtcv$. W.p. $\eps_{\tx{G}}$, randomly generate an adjacency matrix $\tilde{A}$.
    \FOR{$i$ from $1$ to $n$}
        \FOR{$j$ from $1$ to $n$ and $j\not= i$}
            \STATE If $\tilde{A}_{ij} = 0$ set $\mtcn_i^{\tx{(neigh)}} \Let \{q \in \mtcv: \tilde{A}_{iq} = 1\} \cup \{j\}$, and if $\tilde{A}_{ij} = 1$ set $\mtcn_i^{\tx{(neigh)}} \Let \{q \in \mtcv: \tilde{A}_{iq} = 1\} \setminus \{j\}$. Set $\mtcn_i^{\tx{(non)}} \Let \mtcv\setminus \mtcn_i^{\tx{(neigh)}}$. 
            \begin{align*}&([\hat{A}_{\tx{temp}_j}]_{i,:},\hat{\theta}_{\tx{temp}_j}^{(M[k_i(l)],i)})= \tx{Learn}M[k_i(l)]^+ \\
            &([\bm{X}_{\tx{tr}}]_{1:\tilde{T}-1,:},[\bm{X}_{\tx{tr}}]_{2:\tilde{T},i}, \mtcn_i^{\tx{(neigh)}},\mtcn_i^{\tx{(non)}})
            \end{align*}
        \ENDFOR
        \STATE Compute the validation error of $f^{M[k_i(l)]}$ based on $([\hat{A}_{\tx{temp}_j}]_{i,:},\hat{\theta}_{\tx{temp}_j}^{(M[k_i(l)],i)})$, $j \in \mtcv\setminus \{i\}$. Choose the smallest error $p_{i,k_i(l)}(l)$ with respect to $j$, update $[\hat{A}^{(M[k_i(l)])}(l)]_{i,:}$, $\hat{\theta}^{(M[k_i(l)],i)}(l)$ accordingly. Let $p_{i,m}(l) = \infty$, $[\hat{A}^{(M[m])}(l)]_{i,:} = [\hat{A}^{(M[m])}(l-1)]_{i,:}$, and $\hat{\theta}^{(M[m],i)}(l) = \hat{\theta}^{(M[m],i)}(l-1)$ for $m = \{1,2,3,4\}\setminus\{k_i(l)\}$.
    \ENDFOR
    \FOR{$i$ from $1$ to $n$}
        \STATE Find 
        \[\mtcv_i^{(\tx{incst})} \Let \{j \in \mtcv: [\hat{A}^{(M[m_j])}(l)]_{ji} \not = [\hat{A}^{(M[k_i(l)])}(l)]_{ij}\},\]
        where $m_j = \arg\min_{1\le m \le 4} \min_{1\le q\le l} p_{j,m}(q)$.
        \STATE Generate test neighbor sets, $\tilde{\mtcn}_i^{\tx{(neigh,1)}}, \dots, \tilde{\mtcn}_i^{\tx{(neigh},b_\tx{G})}$, according to a budget $b_{\tx{G}} \in \ZZ^+$, such that if $u \in \{j:[\hat{A}^{(M[k_i(l)])}(l)]_{ij} \not=0\}\setminus \mtcv_i^{(\tx{incst})}$ then $u \in \tilde{\mtcn}_i^{\tx{(neigh},q)}$ and if $u \in \{j:[\hat{A}^{(M[k_i(l)])}(l)]_{ij} =0\}\setminus \mtcv_i^{(\tx{incst})}$ then $u \not\in \tilde{\mtcn}_i^{\tx{(neigh},q)}$ for all $1\le q\le b_{\tx{G}}$. Let $\tilde{\mtcn}_i^{\tx{(non},q)} = \mtcv \setminus \tilde{\mtcn}_i^{\tx{(neigh},q)}$.
        \FOR{$q$ from $1$ to $b_{\tx{G}}$}
            \STATE Compute
            \begin{align*}&([\hat{A}_{\tx{temp}_q}]_{i,:},\hat{\theta}_{\tx{temp}_q}^{(M[k_i(l)],i)}) = \tx{Learn}M[k_i(l)]^+\\
            &([\bm{X}_{\tx{tr}}]_{1:\tilde{T}-1,:},[\bm{X}_{\tx{tr}}]_{2:\tilde{T},i}, \tilde{\mtcn}_i^{\tx{(neigh},q)},\tilde{\mtcn}_i^{\tx{(non},q)})
            \end{align*}
        \ENDFOR        
        \STATE Compute the validation error of $f^{M[k_i(l)]}$ based on $([\hat{A}_{\tx{temp}_q}]_{i,:},\hat{\theta}_{\tx{temp}_q}^{(M[k_i(l)],i)})$. Choose the smallest error with respect to $q$ and update $p_{i,k_i(l)}(l)$, $[\hat{A}^{(M[k_i(l)])}(l)]_{i,:}$, $\hat{\theta}^{(M[k_i(l)],i)}(l)$ accordingly, and
        \[[Q(l)]_{i,k_i(l)} = (1 - \alpha)[Q(l-1)]_{i,k_i(l)} - \alpha\log(p_{i,k_i(l)}(l)).\] 
    \ENDFOR
\ENDFOR
\STATE Let $\hat{m}_i \Let \arg\min_{1\le m \le 4} \min_{1\le q\le n_{\tx{iter}}} p_{i,m}(q)$, $\hat{\ModelFori} \Let M[\hat{m}_i]$, $[\hat{A}]_{i,:} = [\hat{A}^{(\hat{\ModelFori})} (n_{\tx{iter}})]_{i,:}$, and $\hat{\theta}^{(\hat{\ModelFori},i)} = \hat{\theta}^{(\hat{\ModelFori},i)}(n_{\tx{iter}})$, $i\in \mtcv$.
\STATE Return $\hat{A}$, $\{\hat{\ModelFori}\}$, and $\{\hat{\theta}^{(\hat{\ModelFori},i)}\}$.
\end{algorithmic}
}
\end{algorithm}

\subsection{Improvement of Learning Algorithm}\label{sec_improv}

In this section, we modify the learning algorithm to exploit the assumptions more efficiently. 

First, note that in Line~11 of Algorithm~\ref{alg_eps} the non-neighbor set is only set to be empty or $\{j\}$. Learn$M[k_i(l)]$ can thus return new unexpected neighbors, possibly worsening the performance of network recovery. To address this issue, we completely fix the neighbor and the non-neighbor sets and run constrained least-squares (Algorithms~\ref{alg_DG+}--\ref{alg_sign+}), rather than Algorithms~\ref{alg_DG}--\ref{alg_sign}. For the learning of signed network, note from~\eqref{eq_M3} that $\bfl^\tp \theta^{(\ModelRepell,i)} = 1$, so another constraint $\bfl^\tp y = 1$ is added.

In addition, we take advantage of the assumption that the underlying graph is undirected to fix possible inconsistent adjacency estimates (that is, for $\hat{A}$ obtained by Algorithm~\ref{alg_eps}, $\hat{A}_{ij} \not= \hat{A}_{ji}$). To this end, at each iteration~$l$ and for each agent~$i$, we collect all estimates $[\hat{A}^{(M[m_j])}(l)]_{ji}$ with $j\in \mtcv$ and $m_j$ the candidate update rule of $j$ with smallest validation error up to and including iteration $l$ (see Algorithm~\ref{alg_eps+}). That is, the estimate of $a_{ji}$ given by the candidate update rule of $j$ with the smallest validation error. Then we compare them with $[\hat{A}^{(M[k_i(l)])}(l)]_{ij}$ at the current iteration to get all $j \in \mtcv$ such that $[\hat{A}^{(M[m_j])}(l)]_{ji} \not = [\hat{A}^{(M[k_i(l)])}(l)]_{ij}$ and collect them in the set $\mtcv_i^{(\tx{incst})}$. The algorithm tests if keeping the edges connecting these agents to $i$ can improve the validation error for the agent~$i$. The detailed procedure is summarized in Lines~15--22 of Algorithm~\ref{alg_eps+}. For notation simplicity and consistency, in Algorithm~\ref{alg_eps+} we denote Learn$\ModelHK^+ \Let$ Learn$\ModelHK$, and set each algorithm $\tx{Learn}\ModelFori^+$ to return $[\hat{A}]_{i,:}$, which has indeed been determined by the algorithm's input $(\mtcn_i^{\tx{(neigh})},\mtcn_i^{\tx{(non})})$.

\begin{figure*}
\centering
    \subfigure[\label{roc10+}$T=10$]{    \includegraphics[width=0.18\linewidth]{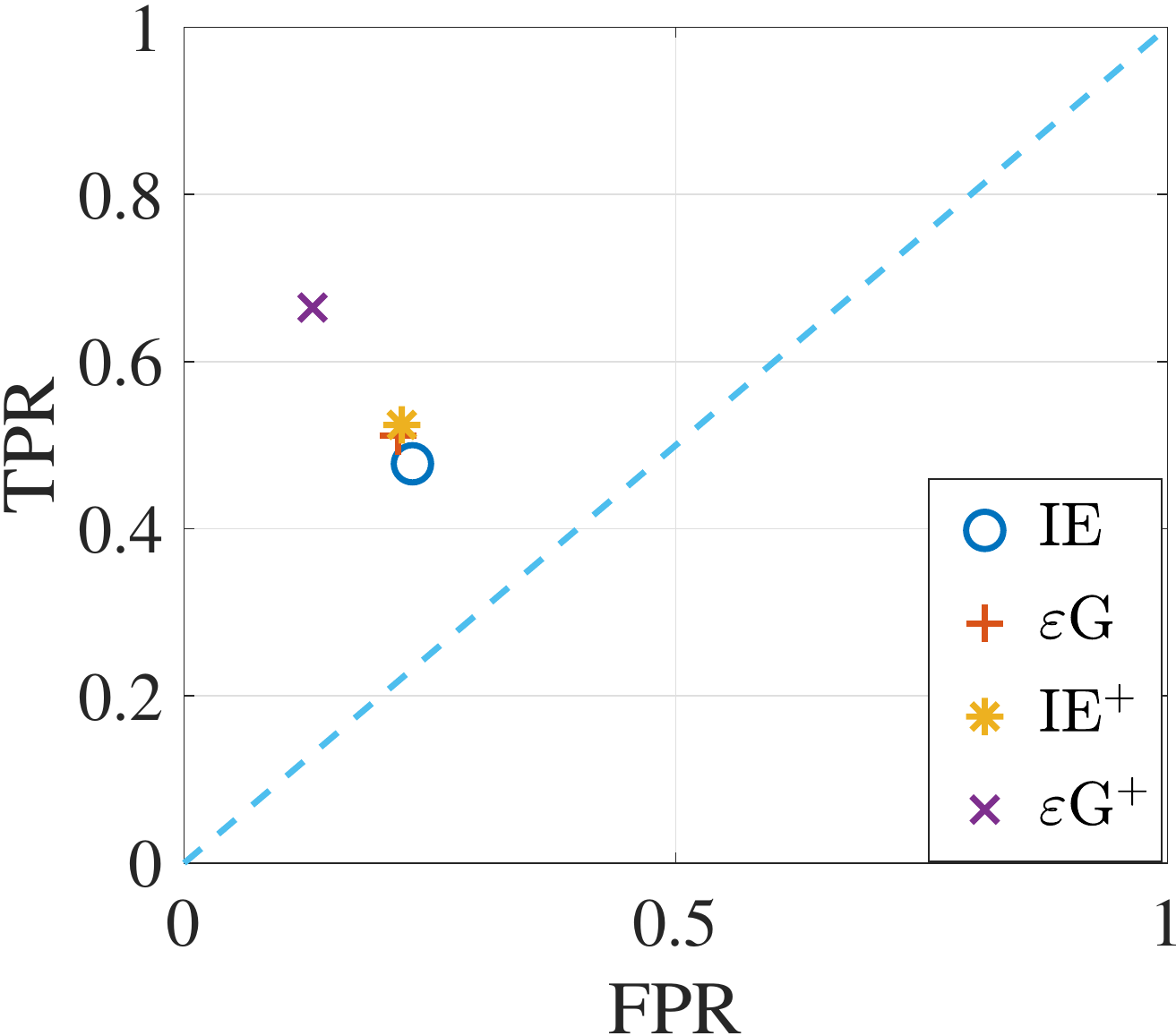}}\quad\quad
    \subfigure[\label{roc15+}$T=15$]{\includegraphics[width=0.18\linewidth]{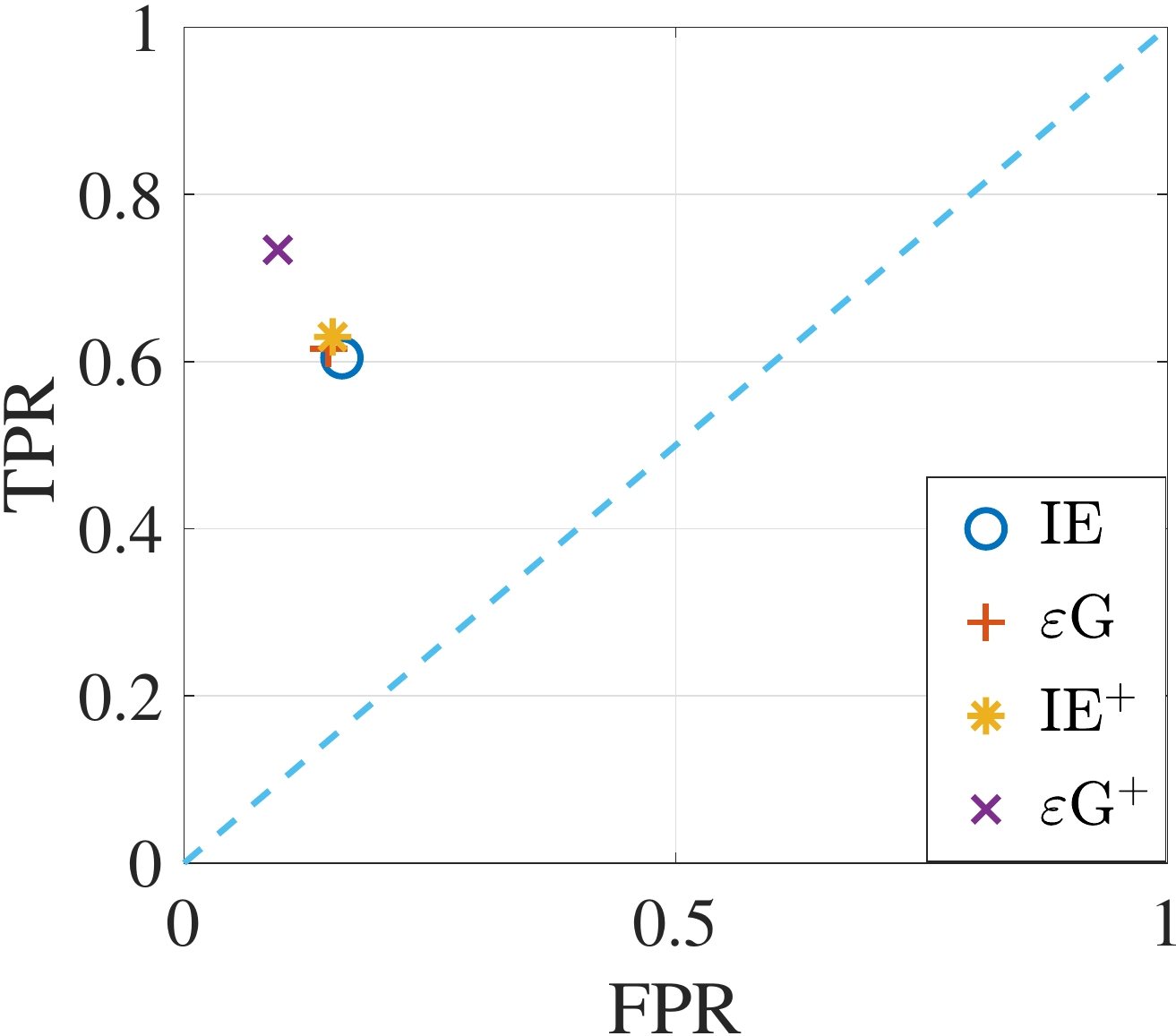} }\quad\quad
    \subfigure[\label{roc20+}$T=20$]{\includegraphics[width=0.18\linewidth]{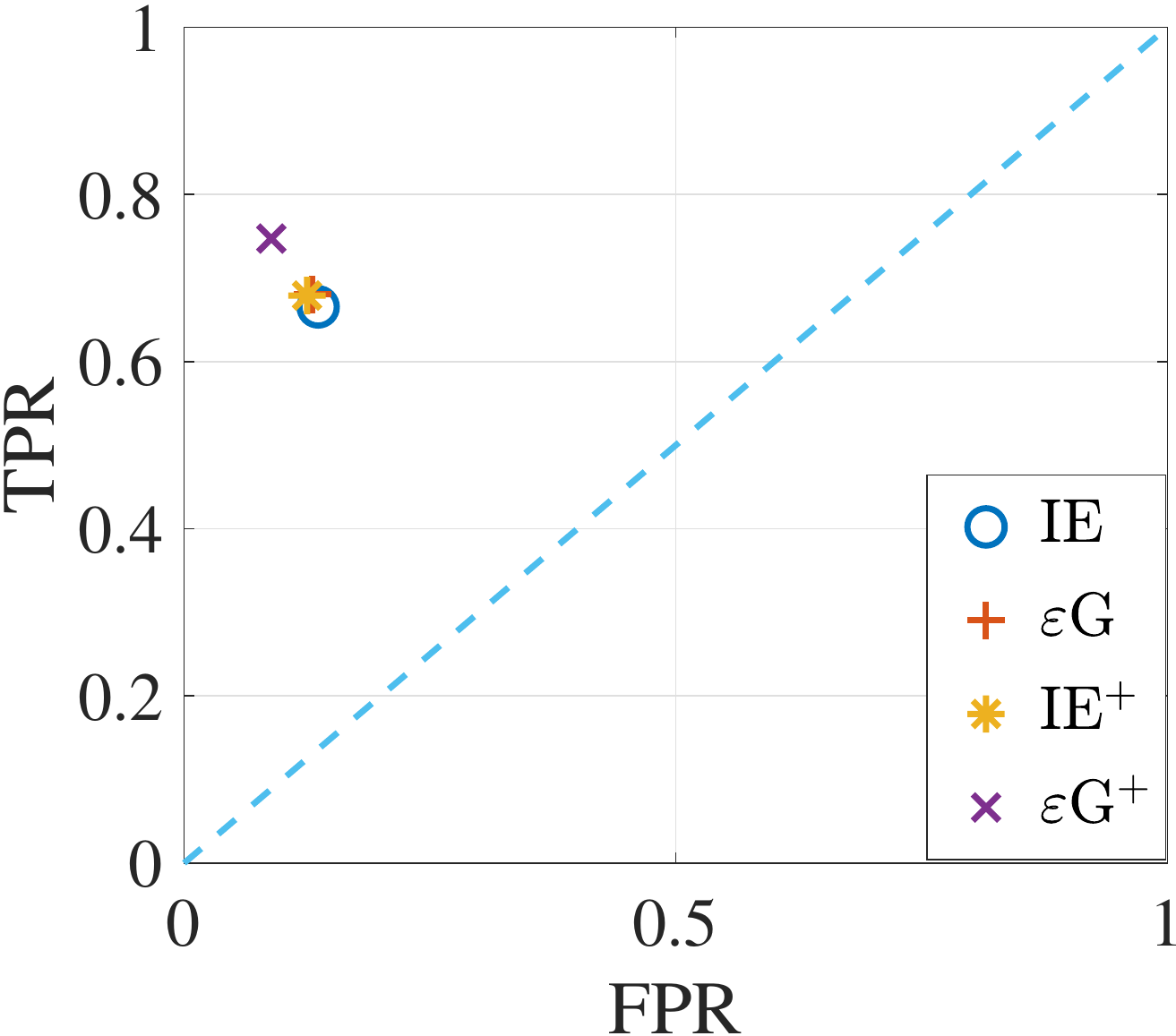} }
    \caption{\label{fig_roc+} Performance of algorithms IE$^+$ and $\eps$G$^+$ for network topology recovery.} 
\end{figure*}

\begin{figure}
\centering 
\includegraphics[width=0.5\linewidth]{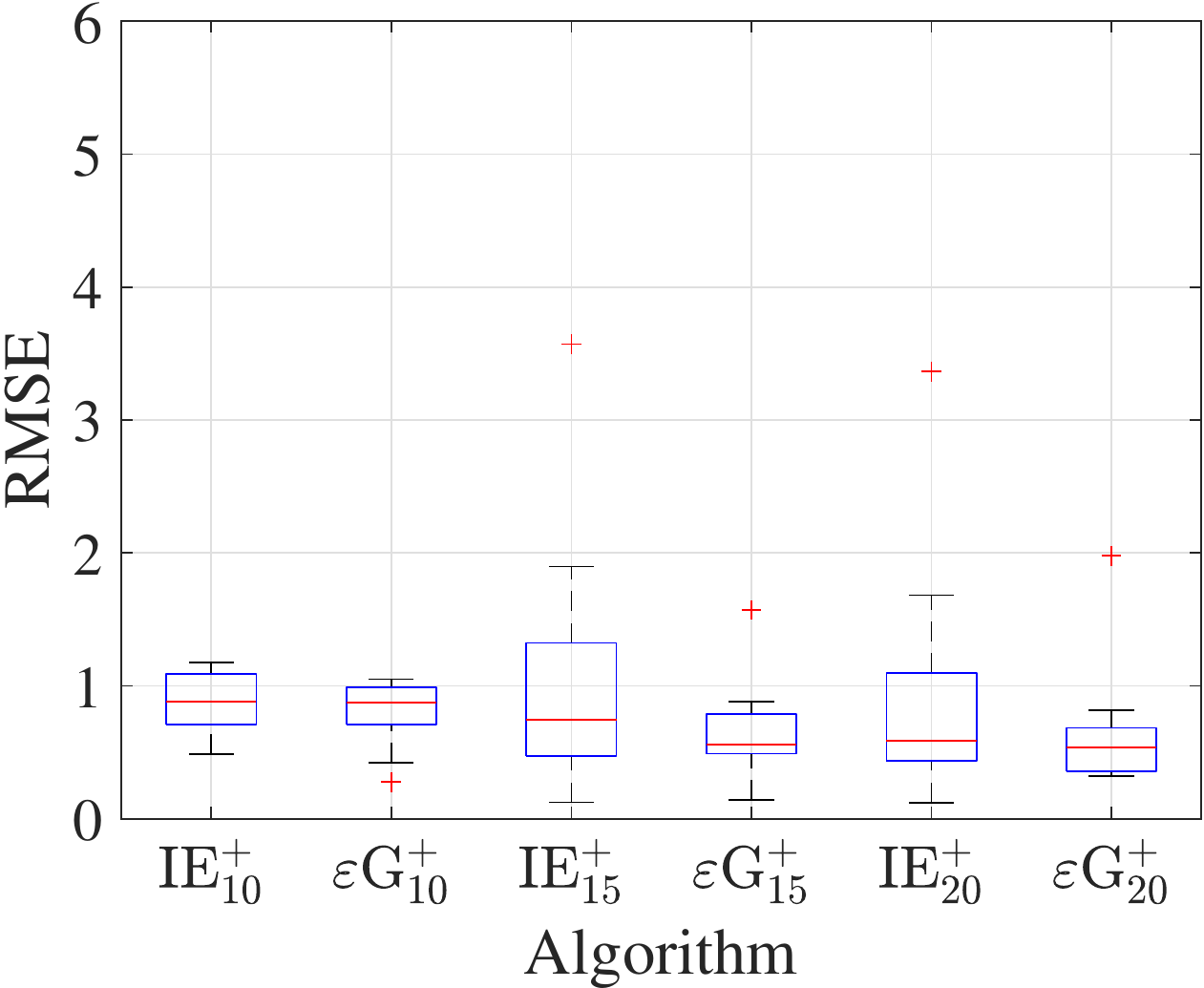}
    \caption{\label{fig_err+} Prediction error of algorithms IE$^+$ and $\eps$G$^+$, where the subscripts indicate the length of trajectories used by the algorithms.} 
\end{figure}

To illustrate the performance of Algorithm~\ref{alg_eps+} (denoted by $\eps$G$^+$), we apply it to the same data set as previously. The algorithm starts with an initial estimate (denoted by IE$^+$), the same to IE except that for $\ModelRepell$ the constraint $\bfl^\tp y = 1$ is added. The hyperparameters of the algorithm are set to be the same as Algorithm~\ref{alg_eps}. Fig.~\ref{fig_roc+} shows that IE$^+$, equipped with the new constraint for $\ModelRepell$, performs better than IE in topology recovery. Moreover, $\eps$G$^+$ improves IE$^+$ significantly in topology recovery because of the introduction of an edge fixing step. Fig.~\ref{fig_err+} illustrates that $\eps$G$^+$ reduces the prediction error of IE$^+$, and that the error of both IE$^+$ and $\eps$G$^+$ decreases with the length of trajectories.

\section{Conclusion}
In this paper, we studied joint learning of topology and dynamics for a mixed opinion dynamics model and proposed a learning algorithm. The proposed algorithm can improve the initial estimates of the network and the update rules, and reduce prediction error. Future work will study improvement and generalization of the proposed algorithm.

\bibliography{ifacconf}             
                                                   







\end{document}